\input epsf

\def\centerwmf#1#2#3{\centerline{\epsfxsize=#1\epsfysize=#2\epsfbox{#3}}}

%\input setwmf.tex

%\input bbb.tex
%nott0515.tex\bigskip

\magnification\magstep1
%\input mssymb.tex
%\input eightpoi.tex
%\font\ftitle=cmbx10 scaled\magstep2
\font\bff =cmcsc10

\def\reals{{{\rm I} \kern -.15em {\rm R} }}
\def\complex{{{\rm I} \kern -.52em {\rm C} }}
\def\nat{{{\rm I} \kern -.15em {\rm N} }}
\def\Z{{{\rm  Z} \kern -.15em \!\!{\rm Z}}}
\def\Q{{{\rm  I} \kern -.53em {\rm Q}}}

\bigskip

\centerline{\bff Non--locality and quantum theory:
                new experimental evidence}
\bigskip
\bigskip
\bigskip

$$\vbox { \settabs  2 \columns
\+ {\bff Luigi Accardi  } &               {\bff Massimo Regoli}  \cr }$$
\bigskip
\bigskip
\centerline{\bff Centro Vito Volterra }
\bigskip\bigskip
\centerline{  Universit\`a degli Studi di Roma ``Tor Vergata'', Roma, Italy }

\centerline{  email: accardi@volterra.mat.uniroma2.it,
              WEB page: http://volterra.mat.uniroma2.it}

\bigskip
\bigskip
\bigskip
\bigskip
\centerline{\bff Abstract}
\bigskip
Starting from the late 60's many experiments have been performed  to
verify the violation Bell's inequality by Einstein--Podolsky--Rosen
(EPR) type correlations.
The idea of these experiments being that: (i) Bell's inequality is a
consequence of locality, hence its experimental violation is an indication
of non locality; (ii) this violation is a typical quantum phenomenon because
any classical system making local choices (either deterministic or random)
will produce correlations satisfying this inequality.

Both statements (i) and (ii) have been criticized by quantum probability on
theoretical grounds (not discussed in the present paper) and the
experiment discussed below has been devised to support these theoretical arguments. We
emphasize that the goal of our experiment is not to reproduce classically the
EPR correlations but to prove that there exist perfectly local classical
dynamical systems violating Bell's inequality.

The conclusions of the present experiment are:
\item{(I)} no contradiction between quantum theory and locality can be
deduced from the violation of Bell's inequality.
\item{(II)} The Copenhagen interpretation of quantum theory becomes
quite reasonable and not metaphyisic if interpreted at the light of the
chameleon effect.
\item{(III)} One can realize quantum entanglement by classical computers.

In our experiment a common source (a computer) generates pseudo--random
points in the unit disk (analogue of singlet pairs) and sends them, one by
one, to two spatially separated experimenters. Each of them, given such a
point, asks an independent local binary question (analogue of a local
measurement of spin in one direction). The answers, coded as
$\pm1$, are obtained by locally calculating the value of a function depending
only on the local question and on the common points in the disk.
The algorithm to calculate these answers is completely local (no
computer knows which question has been asked to the other one) but devised
so to guarantee entanglement: if the same question is posed, opposite
answers will be given.
The questions are parametrized by vectors in the unit circle and we prove that
three such vectors can be chosen so that the corresponding correlations
violate Bell's inequality. In section (7) we prove that
our experiment also provides a classical analogue of the type of logical
(i.e. independent of statistics) incompatibilities pointed out
by Greenberger, Horne and Zeilinger.

\vfill\eject

\beginsection{\bff (1) Introduction}

\bigskip
The experiment described in the present paper has been performed in the
attempt to clarify a question that has accompanied
quantum theory since its origins:
{\it is it true that the theory of relativity, quantum mechanics and a
realistic interpretation of natural phenomena are mutually contradictory}?

In the past 30 years the answer to this question, accepted by the
majority of physicists has been: {\it yes they are and this can be proved
by theory and confirmed by experiment}. Furthermore, since the
experimentally confirmed EPR type correlations are {\it $\ldots $
necessarily nonlocal in character $\ldots $ } [GrHoZe93], the
experiments also solve the contradiction in favor of quantum theory by
showing that the basic pillar of relativity theory is violated in
nature. Usually this statement is tempered by the clause that quantum
nonlocality cannot be used to send superluminal signals. This amounts to
downgrade the locality principle from a law of nature (no action at
distance exists) to a principle of telecommunications theory (the
existing actions at distance cannot be used to build up a superluminal
television). Of course, if this downgrade is the only reasonable way to
interpret the experimental data, we have to accept it. However, given
the relevance of the issue, the question whether it effectively is the only
reasonable way to interpret the experimental data, naturally arises.
In several papers starting from [Ac81a], (cf. [Ac97] for a general
presentation, [Ac99], [AcRe99a] for more recent results)
the arguments, relating Bell's inequality to locality, have been
criticised on a theoretical ground. In the present paper we will
substantiate these theoretical arguments with an experiment.

Before describing our experiment let us briefly review the usual proof of
the contradiction between relativity, quantum mechanics and realism, which
is based on a combination of Einstein, Podolsky,
Rosen (EPR) type arguments with Bell's inequality and goes as follows.
In the EPR type experiments a source emits pairs of systems with the
following properties:
\item{i)} the members of each pair are distinguishable after separation and
we denote them $1$ and $2$. After emission the two become spatially
separated, say: $1$ goes to the left, $2$ goes to the right.
%\vskip5truecm

\item{ii)} for each member $j=1,2$ of each pair we can measure a family
of observables (spin)
$$\hat S^{(j)}_a\quad;\qquad j=1,2$$
parametrized by an index set $T$ ($a\in T$). To fix the ideas let us say
that $T$ is the unit circle so that each $a$ is a unit vector in the plane.

\item{iii)} each observable $\hat S^{(j)}_a$ $(j=1,2;a\in T)$ can take only
two values: $\pm1$ and, if $a\not=b$, the two observables
$$\hat S^{(j)}_a\ ;\qquad \hat S^{(j)}_b \qquad ; \qquad j=1,2 $$
cannot be simultaneously measured on the same system
\item{iv)} The {\it singlet condition} is satisfied, i.e. if the same
observable is measured on both particles, then the results are opposite.
\bigskip
According to EPR the values of the spins must be {\it pre--determined}
otherwise, if one assumes that {\it these values are created} by a choice of
nature at the act of measurement, the only way to explain the strict
validity of the singlet condition would be to postulate a mechanism of
instantaneous action at distance through which a particle {\it instantaneously
knows} which choice nature is going to do on its distant partner so that the
two choices can be {\it matched} into the singlet law. The pre--determination
of the result is called a {\it realism condition}. Thus the negation of the
realism condition in the sense of EPR, would imply a nonlocality
effect. We anticipate that, in our experiment, a crucial role will be played
by a distinction between the EPR realism , or {\it ballot box realism}, on
which the whole of classical statistics is based, from what might be called
the {\it chameleon realism } which seems to be more appropriate when dealing
with quantum systems.

Before explaining this distinction let us shortly outline the role of
Bell's inequality: {\it if the values of the spins must
be pre--determined, then there must be a hidden parameter $\lambda$
which determines these values}. In other words the spin variables must be
functions $S^{(j)}_a$ of this hidden parameter $\lambda$:
$$\hat S^{(j)}_a = S^{(j)}_a (\lambda ) \qquad ; \qquad (j=1,2;a\in T)$$
and the statistical fluctuations of the spins reflect a statistical
distribution of this unknown parameter. The nature of the hidden
parameter and of its statistical distribution are both unknown to us,
but Bell proves that, whatever this nature may be, the correlations
$$ \langle \hat S^{(1)}_a \hat S^{(2)}_b\rangle $$
must satisfy a certain inequality that now brings his name.

The quantity of experimental data accumulated in EPR type experiments is now
quite remarkable and the common consensus on the interpretation of these data
is that they exhibit a violation of Bell's inequality. A survey of
the situation up to 1988 was published in {\it Nature} by
J. Maddox [Mdx88] and a multiplicity of results in the past 12
years have further confirmed the experimental violation of this
inequality by a variety of quantum systems.

According to Bell, given the realism condition (as defined above), the {\it
vital assumption} in the deduction of his inequality is a locality condition:

{\it ... The vital assumption [2] [i.e. the EPR paper [EPR35]] is that
the result $B$ for particle 2 does not depend on the setting $\vec a$ of
the magnet for particle 1, nor $A$ on} $\vec b\dots$ [Be64]

A theory which satisfies both assumptions is called a {\it local realistic
theory}. With this terminology Bell's result is often reformulated in the form:
{\it a theory that violates Bell's inequality must be either
non--local or non--realistic}.

Since, by the original EPR argument, a non--realistic theory is
necessarily non--local, it follows that in any case the violation of Bell's
inequality implies non--locality.
This is Bell's original thesis and, even if balanced by a multiplicity of
subtle differences on minor points, from the huge literature now available
on this topic (cf. for example [Red87], [Be87], [CuMcM89], [Maud94], ...),
a substantial consensus emerges with its main points which one may
summarize in the scheme:

\item{--} locality implies Bell's inequality

\item{--} quantum mechanical (EPR type) experimental correlations violate
          Bell's inequality

\item{--}  therefore quantum mechanics is non--local

\item{--}  furthermore, no classical system can
violate Bell's inequality by local choices.
\bigskip
Therefore if we can construct a classical
system (even deteministic and macroscopic) which, by means of purely
local choices produces a set of statistical correlations violating
Bell's inequality, this will be an experimental proof of the fact that
\item{(i)} locality does not imply Bell's inequality
\item{(ii)} the experimental violation of Bell's inequality achieved
by the quantum mechanical (EPR type) correlations is not an indication
of non--locality.
\item{(iii)} the experimental violation of Bell's inequality by purely
local experiments is not a typically quantum phenomenon
\bigskip
It is since [Ac81a] known that, if $\hat S^{(1)}_a$, $\hat S^{(2)}_b,\dots$
are functions defined
on the same probability space then the inequality (4) below must be fulfilled
even if we postulate a strong non locality condition (e.g. that
$\hat S^{(1)}_a$ {\bf depends} on the values assumed by $\hat S^{(2)}_b$, cf. [Ac81]).

Therefore, if we want to violate (4) we have to find a local mechanism of
generation of the answers so that each pair
$(\hat S^{(1)}_a,\hat S^{(2)}_b),\dots$ have a joint
probability distribution, but this is not true for triples.
Our goal in this paper is to construct such a system.

The goal of the present experiment is not to build a hidden variable model
for the quantum mechanical singlet correlations (although a variant of it
may play this role), but to construct a classical system showing that the
three conditions of pre--determination, locality and singlet condition are
not incompatible with a violation of Bell's inequality.
Consequently the violation of these inequalities in the EPR--type
experiment cannot be interpreted as an incompatibility of the three
above mentioned conditions with quantum theory.
\bigskip

\beginsection{\bff (2) Description of the experiment}

\bigskip
Our experiment describes a classical situation analogous,
but not identical, to the one which takes place in the EPR type experiments.
The classical system considered is made of three computers:

\item{(2.1)}  one, playing the role of the source of the singlet pairs,
will produce pseudo--random points in the unit disk
$$p_1,p_2,\dots,p_N\eqno(1)$$
Each point $p_j$ plays the role of a singlet pair, therefore in the
following we shall speak indifferently of {\it the point } $p_j$ or of
{\it the singlet pair } $p_j$.
The choice of the pseudo--random generator is not relevant for the results,
provided the algorithm has a reasonably good performance. The algorithm
we use has been taken from [PrTe93]. An order of $N=50.000$
points is already sufficient to obtain acceptable results. Increasing
the number of points gives a higher precision in the computation of the
areas, but does not change the order of magnitude of the violation of
the inequality. This also shows that the violation we obtain cannot be
attributed to round--off errors in the measurements of the areas.
\item{(2.2)}  the other two computers play the role of the two measurement
apparata. The three computers are separated (e.g. they may be in
different rooms or different countries, $\ldots$).
\item{(2.3)}  The role of the experimentalists will be played by two persons,
one for each of the "measuring computers". Given a point $p_j$ (a singlet
pair), operator $1$ makes a local independent choice of a unit
vector $a$ in the plane (analogue of the direction of the magnetic field)
and activates a programme in computer $1$ which computes the value of a
function $S^{(1)}_a(p_j)$, depending only on the local choice $a$ and on the
point (singlet pair) $p_j$. Similarly operator $2$ computes the value
$S^{(2)}_b(p_j)$, depending only on her own local choice $b$, but (as in
all EPR experiments) on the same singlet pair $p_j$.
We shall say that each operator {\it asks a (local) question to the system
(computer)}
\item{(2.4)} the values of the functions $S^{(1)}_a$, $S^{(2)}_b$
(answers) can only be $+1$ or $-1$.
\item{--} A priori, for any given point (singlet pair) $p_j$,
experimenter $1$ (resp. $2$) can choose among infinitely many
questions to be asked to system $1$ (resp. $2$).
However only one question at a time can be asked to any single system.

\item{(2.5)} The calculation of the values $S^{(1)}_a(p_j)$, $S^{(2)}_b(p_j)$
is purely local, however the programme has been devised so that if, by
chance, the same question is asked to both systems then they will give
opposite answers (singlet condition).

\item{(2.6)} Finally we introduce, in our classical model, a strong form
of the disturbance effect of a measurement of an observable on the other
observables of the same system, incompatible with the given one, by requiring
that, if the observable $\hat S_x$ is measured, then both types of particles
instantaneously change the value of all the other observables from
$\hat S_y\ (y\not=x)$ to $-\hat S_y$ (if I measure the weight of a chameleon
in a closed box, its color need not to be the same I would have found if I
would have measured it on a leaf). This additional prescription plays a
role in the deduction of the Bell inequality (section (5)), but is not
necessary for the Zeilinger--Harne--Greenberger type contradiction of
section (7).
\bigskip
It is clear from our rules that the answers are:
\bigskip
\item{i)} {\it pre--determined\/}, i.e. each system of the pair knows a priori
what his/her answer will be if any question, determined by an input
point $p_j$ and a local choice $a$ or $b$, will be asked.
\item{ii)} {\it local\/}, i.e. if, for a given point $p_j$,
operator $1$ makes the local choice $a$, the answer of system $1$ depends
{\bf only} on $p_j$ and $a$. In particular it {\it does not
depend\/} on which question $b$ has been asked to system $2$. The same
is true exchanging, in the above statement, the roles of $1$ and $2$.
\item{iii)} {\it entangled\/}, i.e. the singlet condition is satisfied.
\bigskip
We want to study the statistics of these answers. In order to do so we
have to ask the same pair of questions to many, say $N$, pairs:
$p_1,\dots,p_N$. In particular, denoting
$$S^{(\nu)}_a(p_{j})\ ;\quad\nu=1,2\ ;\quad j=1,\dots,N\eqno(2)$$
the answer given by system $\nu(=1,2)$ of the $j$--th pair to the local
questions $S^{(\nu)}_a$, we can define the {\it empirical correlation\/}
for any pair $(S^{(1)}_a,S^{(2)}_b)$ of local questions
$$\langle S^{(1)}_a S^{(2)}_b\rangle={1\over N}\,\sum^N_{j=1}
S^{(1)}_a(p_{j})S^{(2)}_b(p_{j})\eqno(3)$$

The question we want to answer with our experiment is the following:

\noindent{\it can the members of each pair $p_j$ make an agreement on
their answers so that the inequality\/}
$$|\langle S^{(1)}_aS^{(2)}_b\rangle-\langle S^{(1)}_cS^{(2)}_b\rangle
|\leq1+\langle S^{(1)}_aS^{(2)}_c\rangle\eqno(4)$$
{\it is violated for some choices of the indices $a$, $b$, $c$?\/}

According to Bell's analysis, if the inequality (4) is violated, then at
least one of the conditions (i), (ii), (iii) must be violated (cf.
section 5 for further discussion of this point). Our
experiment contradicts this conclusion.
\bigskip\medskip

\noindent{\bff (3) Intuitive interpretation of the experiment}

\bigskip
The present experiment describes an ensemble of pairs of classical particles
which can interact with a vector observable $\vec B$, that we call
{\it magnetic field\/}.
We assume that there are two types of particles, type I (left handed) and type
II (right handed) and that each pair contains exactly one particle of
type I and one particle of type II.
% Syntax:  \centerwmf{<width>}{<height>}{<path+filename>}
%     Requires "\input setwmf" at the beginning of your file.
% Optional:  <path> (use / instead of \), specifies path of TeX file if not supplied.
% Example:  \centerwmf{3in}{2in}{c:/mysubdir/mypic.wmf}

%\centerwmf{7.53cm}{5.79cm}{fig2.wmf}
%\centerwmf{11.57cm}{5.51cm}{fig2e3.wmf}
The pairs decay, i.e. split, sequentially, one after the other, and we
suppose that left handed particles always go on the left  and right handed
particles always go on the right, so that after a short time they become
spatially separated. On the path of each type of particles there is an
experimentalist and the two have coordinated their space--time reference
frames so that it makes sense to say that they make simultaneous
measurements and that they orient a physical vector quantity in a common
direction. Without loss of generality we can assume that:
\item{--} all the experiments take place in the same plane
\item{--} in this plane we have fixed an oriented reference frame $e=(e_1,e_2)$
\item{--} this reference frame can be parallel transported without holonomy to
          different points of the (affine) plane (Figure (1)).

%
%\vskip6truecm
%
\centerwmf{7.53cm}{5.79cm}{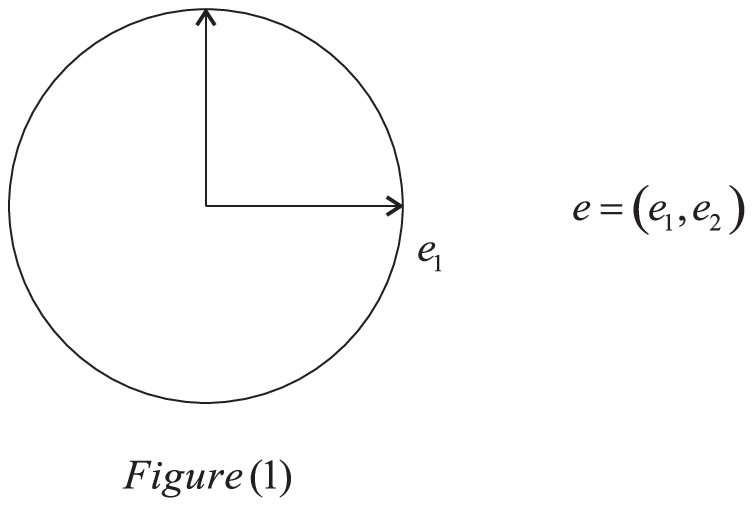}
In particular, with respect to this frame, we can speak of upper and lower
half--plane and of counterclockwise or clockwise measurement of the angles
(Figures (2) and (3)). Moreover for any point $c$ in the disk,we denote
$Rc$ its reflection with respect to the origin (Figure (3))

%\vskip6truecm

\centerwmf{11.57cm}{4.88cm}{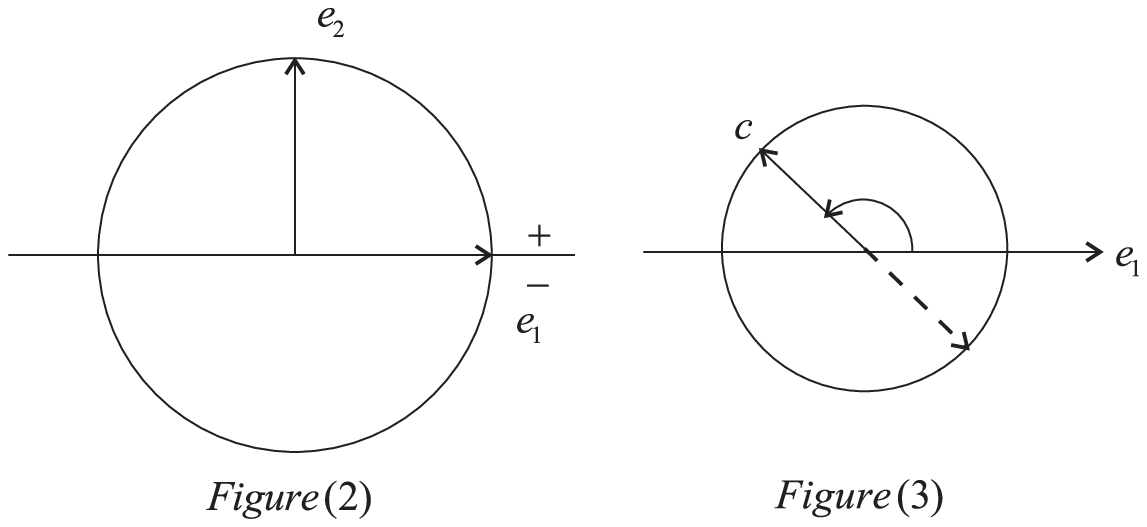}
Denote $D$ the unit disk in the plane:
$$D:=\{p=(x,y)\in\reals^2:x^2+y^2\leq1\}$$

$\partial D$ the unit circle

$D_+=$ (resp. $\partial D_+$) the upper semi--disk (resp. semi--circle)

$D_-=$ (resp. $\partial D_-$) the lower semi--disk (resp. semi--circle)

\noindent (cf. Figure (2)).
To both types of particles we associate a family of observables (that we
call {\it spin\/})
pa\-ra\-me\-tri\-zed by the points on the unit circle in the plane
$$\{\hat S_a:a\in\partial D\}$$
and with values $\pm1$ in appropriate units. The two types of particles differ
because a magnetic field of unit strength in direction $a\in\partial D$ splits
a beam of particles of type I according to the values of the observable $\hat S_a$,
and a beam of particles of type II according to the values of the
observable $\hat S_{Ra}$. More precisely: if $a$ is any unit vector in the upper
half plane (which includes the vector $e_1$, but not the vector $Re_1$) then,
under the action of a magnetic field $B(a)$ in direction $a$, a particle
of type I will deviate in the upward direction if $\hat S_a=+1$, in the downward
direction if $\hat S_a=-1$
$$\hbox{Type I }\to B(a)\to  \cases{ \uparrow, & if $\hat S_a=+$  \cr
                                   \downarrow, & if $\hat S_a=-$  \cr}\eqno(1)$$
where the first arrow means {\it input\/} and the second one denotes the
effect of the interaction with the magnetic field. Under the action of
the same magnetic field $B(a)$, a particle of type II will
deviate in the upward direction if $\hat S_{Ra}=+1$, in the downward direction
if $\hat S_{Ra}=-1$
$$\hbox{Type II }\to B(a)\to  \cases{\uparrow, & if $\hat S_{Ra}=+$ \cr
                                   \downarrow, & if $\hat S_{Ra}=-$  \cr}\eqno(2)$$
For $a$ as above, a magnetic field in direction $Ra$ will have the symmetric
effect:
$$\hbox{Type I }\to B(Ra)\to  \cases{ \uparrow, & if $\hat S_{Ra}=-$  \cr
                                   \downarrow, & if $\hat S_{Ra}=+$  \cr}\eqno(3)$$
$$\hbox{Type II }\to B(Ra)\to  \cases{\uparrow, & if $\hat S_{a}=-$ \cr
                                   \downarrow, & if $\hat S_{a}=+$  \cr}\eqno(4)$$
i.e. particles of type I will go up if $\hat S_{Ra}=-1$, down if $\hat S_{Ra}=+1$ and
particles of type II will go up if $\hat S_{a}=-1$, down if $\hat S_{a}=+1$

In this sense the observable $\hat S_a$ can be understood as measuring the
response of a particle (either of Type or of I Type II) to a magnetic field
in direction $a$, {\it being understood that no other interaction has to take
place with the same particle in the same time}. In fact what we measure
is only this response and the {\it values} $\pm 1$ are simply a
conventional way to code these two mutually exclusive physical alternatives.

From this definition it is clear that for any two different vectors
$a, b\in\partial D$ the corresponding observables $\hat S_a$, $\hat S_b$ are
incompatible because it is impossible to place two different magnetic fields
$B(a),B(b)$ in the same point and, even if it were possible, a single
particle cannot interact {\it only with} $B(a)$ and, at the same time,
{\it only with} $B(b)$. A biological analogue of this situation would be
when $\hat S_a$ denotes the color of a chameleon on a leaf and $\hat S_b$ its color on a
log: it is self--contradictory to say that the color {\it is simultaneously
measured only on a leaf and only on a log}. This expression of logical
impossibility has to be distinguished from the usual statement of the
Heisenberg principle, expressing the {\it physical} (but by no means logical)
impossibility of simultaneously measuring, with arbitrary precision and on
the same system, position and momentum.
\bigskip
\noindent{\bff Remark}. In standard quantum mechanics it is well known that the
spin in direction $a$ is a pseudo--scalar quantity, i.e. under proper
rotations it transforms as a scalar but, under parity, according to the
rule ([Sak85], chap. 4)
$$S_{Ra}= -S_{a} $$
and obviously the two hermitean matrices $S_{a}$ and $-S_{a} $ commute.
However the above considerations on the impossibility, for a single
particle, to interact simultaneously with two magnetic fields oriented
in opposite directions and placed in the same point also apply to spin
variables and show that some care is needed when identifying the
simultaneous physical measurability of two observables with the
commutativity of the associated operators, at least when pseudo--scalar
quantities are involved.
\bigskip
On two different particles any pair of observables can be
simltaneously measured. Accordingly
an ordered pair $(B(a),B(b))$ of magnetic fields in directions
$a, b\in\partial D_+$ in the upper half plane will describe a simultaneous
measurement of observable $\hat S_a$, for particle I, and of observable
$\hat S_{Rb}$, for particle II, of the pair. We will also use the notation
$(S^{(1)}_{a},S^{(2)}_{b})$ for such a measurement.

Given that the two types of particles react differently to the same
apparatus, we have to distinguish between the two statements:

\item{(i)} the same measurement operation is performed on both particle
$1$ and particle $2$

\item{(ii)} the same observable is measured on both particle $1$ and
particle $2$

For example, case (i) corresponds to use the same magnetic field $B(a)$
on both particles of the pair. In this case we know that we are measuring
observable $\hat S_{a}$ on particle $1$ and observable $\hat S_{Ra}$ on particle $2$.

Case (ii) corresponds to use the magnetic field $B(a)$ for particle $1$ and
the magnetic field $B(Ra)$ for particle $2$. In this case we know that we
are measuring the same observable $\hat S_{a}$ on both particles.

With these premises we will discuss the statistics, over an esemble of
pairs, of measurements of pairs of observables of the form
$(S^{(1)}_a,S^{(2)}_b)$.
Since we are interested in the Bell's inequality we will consider three
experiments of the form
$$(B(a),B(b))\quad,\quad (B(c),B(b))\quad,\quad (B(a),B(c))\eqno(5)$$
corresponding to the measurements of the pairs of observables
$$(S^{(1)}_a,S^{(2)}_{b}) \quad , \quad (S^{(1)}_c,S^{(2)}_{b}) \quad , \quad
(S^{(1)}_a,S^{(2)}_{c})\eqno(6)$$
which, in terms of the original $\hat S$--observables can be written
$$(\hat S_a,\hat S_{Rb})\quad , \quad  (\hat S_c,\hat S_{Rb})\quad , \quad  (\hat S_a,\hat S_{Rc})\eqno(7)$$
Notice that we have {\it three\/} experiments, corresponding to the three
magnetic fields
$$B(a),\quad B(b),\quad B(c)\eqno(8)$$
but the observables involved cannot be reduced to three in fact, as made
explicit by the notation (4) they are {\it four\/}, i.e.
$$\hat S_a,\ \hat S_{Rb},\ \hat S_c,\ \hat S_{Rc}\eqno(9)$$
due to the different dynamical reaction of particle I and particle II to the
same magnetic field $B(c)$. This is an example, in a particularly simple
and idealized classical situation, of the {\it chameleon effect} (the
interaction Hamiltonian of a system with an apparatus, hence the dynamical
evolution of the system , depends on the observable one wants to measure)
described, for example, in [Ac99], [AcRe99a].
\bigskip

\beginsection{\bff (4) Prescriptions for the local answers}

\bigskip
The state space of our classical particles is the unit disk $D$
and, for $c$ in the unit circle, we denote $S^{(j)}_c$ the response of
particle $j$ to the magnetic field $B(c)$ as described in section (3).
We will describe these responses by means of functions
$S_x:p\in D\to S_x(p)=\pm1$, parametrized by points $p$ in the unit circle.
Throughout this section, the symbol $S_x$ shall denote such a function
{\bf and should not be confused with the observables $\hat S_x$ of Section (3)}:
for example $S^{(2)}_x$ measures the observabales $\hat S_{Rx}$, but is
calculated using the function $-S_{Rx}(p)=S_x(p)$. The values of the
observables $S^{(j)}_c$ are distributed according to the rule described in
Figures (4) and (5):

\centerwmf{15.8cm}{6cm}{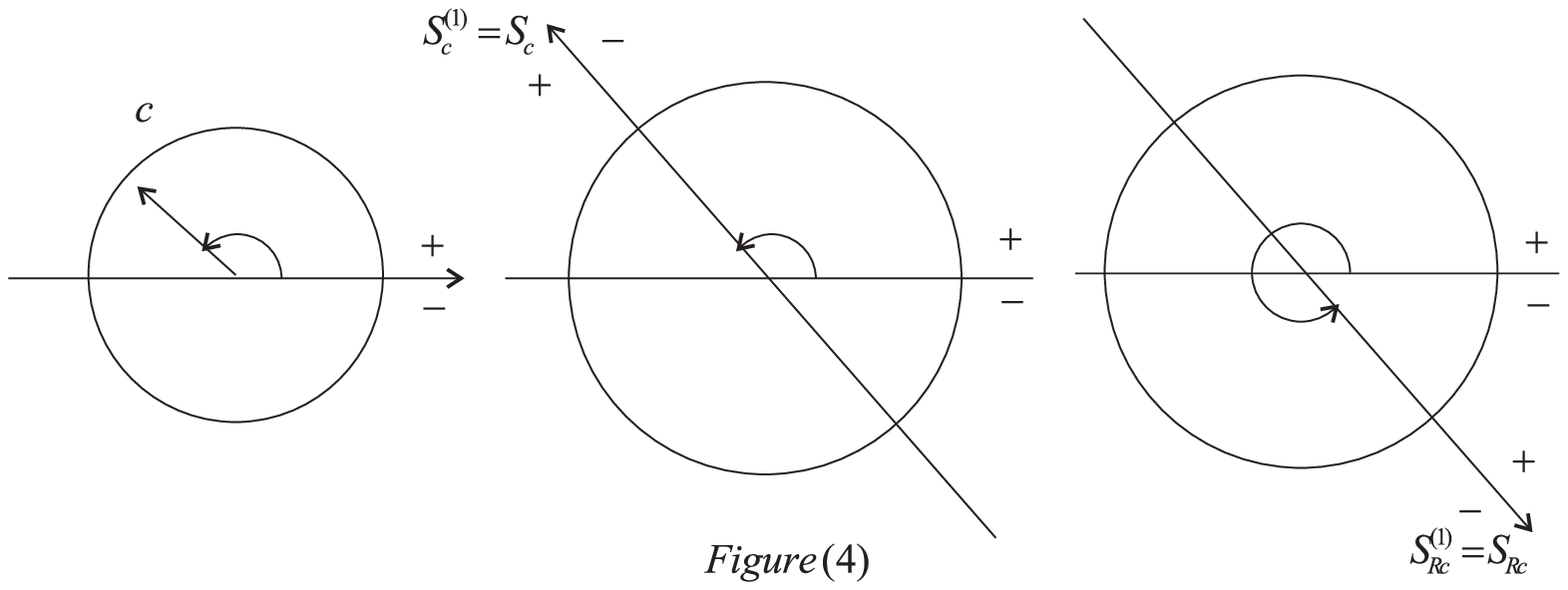}
\centerwmf{15.8cm}{6.1cm}{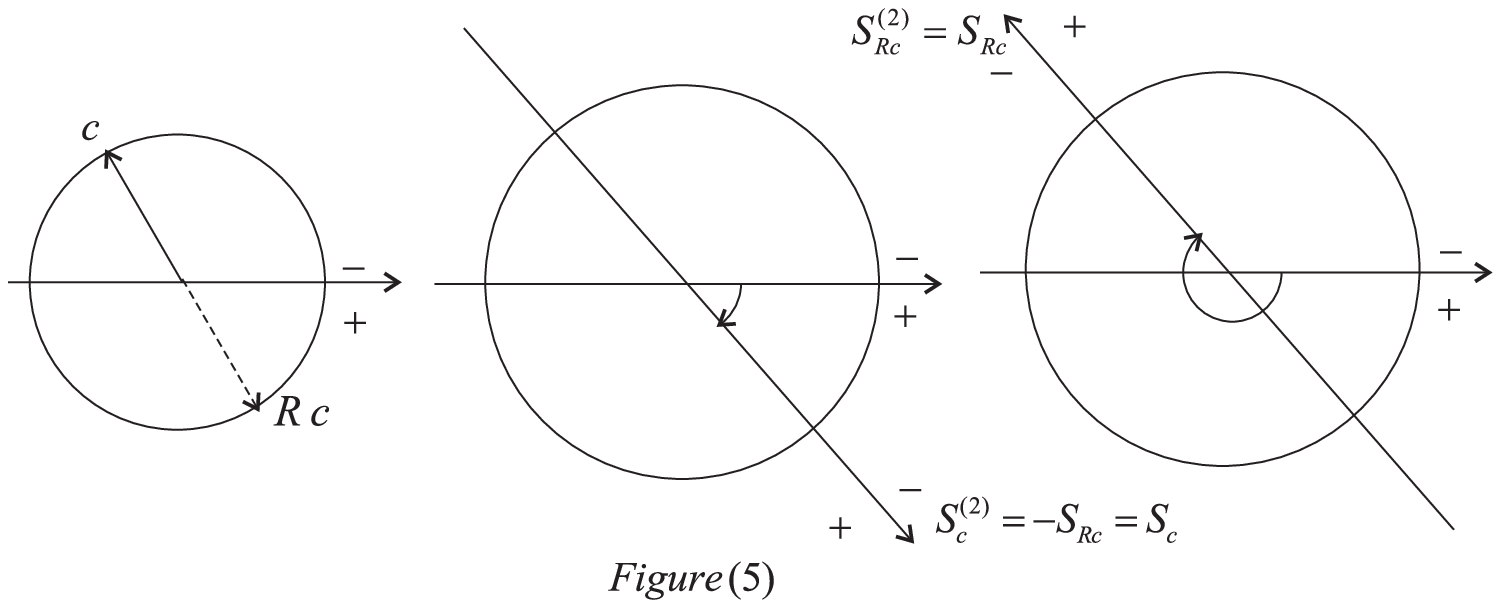}
%\vskip7truecm

%\vskip7truecm

Notice that the two rules give exactly the same function for $c=b$, i.e.
$$S^{(1)}_c=S_c= - S_{Rc}= S^{(2)}_c\eqno(1)$$
As explained in the previous section, this corresponds to the
measurement of $S_c$ on particle $1$ and of  $S_{Rc}$ on particle $2$
while, to measure $S_c$ on both particles we have to place a magnetic
field oriented in direction $c$ for particle $1$ and in direction $Rc$
for particle $2$. This gives, according to (1):
$$S^{(2)}_{Rc} = - S_{R^2c}= - S_{c}=-S^{(1)}_c\eqno(2)$$
which is the singlet law.
\bigskip
The programme to calculate the {\it hidden functions} $S_a$ is the
following. For $a\in \partial D_+$, making a
angle $\alpha \in [0,\pi)$ (measured counterclockwise) with the $x$--axis
(cf. Figure (6)),
%\vskip5truecm
\centerwmf{7.77cm}{6.46cm}{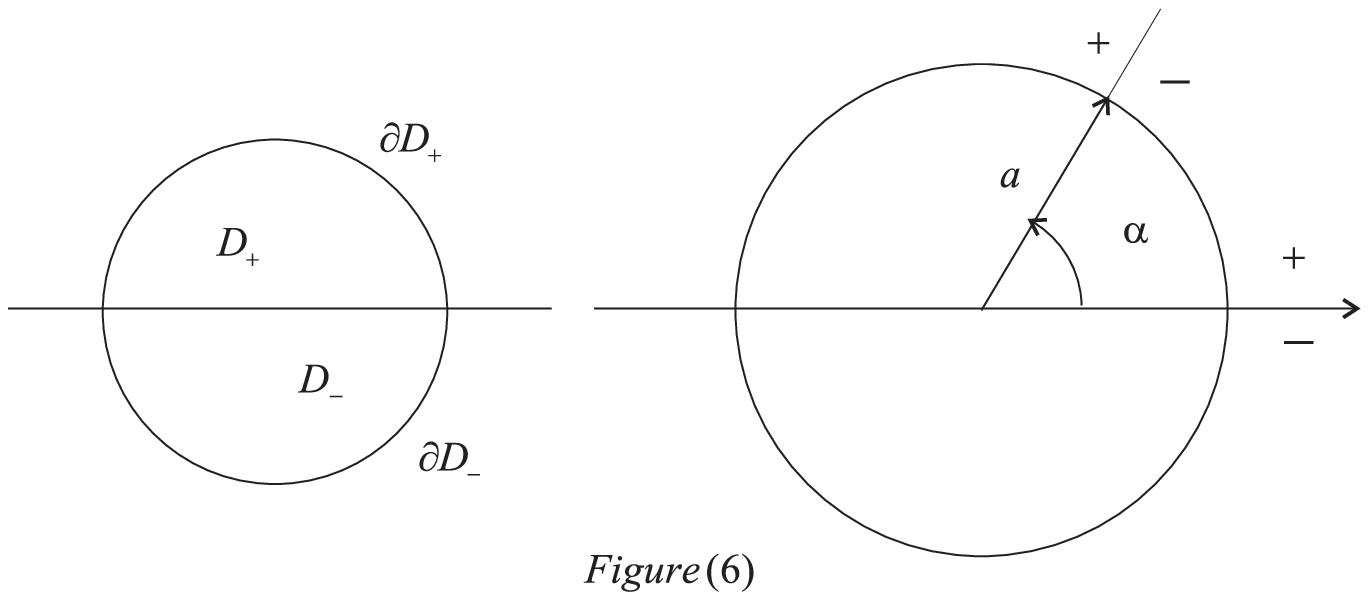}

the calculation of the value of $S_a=:S^{(1)}_a$
in the point $p$ of the disk $D$ is done as follows:
\item{--} one rotates the point $p$, counterclockwise, of an angle
$\alpha$
%%%\centerwmf{5.6cm}{5.68cm}{fig8.wmf}
$$R_{-\alpha}p\qquad ;\qquad R_{-\alpha}=\pmatrix{
\cos\alpha&\sin\alpha\cr
-\sin\alpha&\cos\alpha\cr}\eqno(4)$$

\centerwmf{5.6cm}{5.68cm}{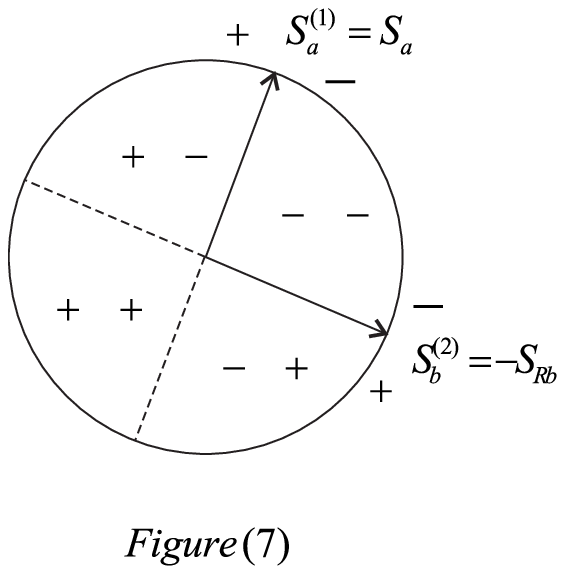}

\item{--} one chooses the value
$$S_a(p)=\pm 1\quad\hbox{if}\quad R_{-\alpha}p\in D_\pm \eqno(5)$$
\item{--} The value of $S_{Ra}(p)$ is determined by the prescription
$$S_{Ra}:=-S_a\eqno(6)$$
Given (4) these prescriptions uniquely determine all pairs
$(S^{(1)}_a, S^{(1)}_b )$.
The statistics of a pair $(S^{(1)}_a, S^{(1)}_b )$
is uniquely determined by the joint
probabilities of concordance, $(+,+)$, $(-,-)$ and discordance, $(-,+)$,
$(+,-)$, which are proportional to the areas of the regions marked in Figure (7).
%\vskip6truecm

\beginsection{(5) The chameleon effect}

\bigskip
In this section we show that, if one pretends to apply to our experiment
the same type of arguments which are applied to the EPR type experiments,
then one arrives to the conclusion that Bell's inequality should be
satisfied and this leads to a contradiction with the experimental data.

Recall that the chameleon effect means that {\it the dynamical evolution
of a system depends on the observable $A$ one is going to measure\/}.
In particular, if we measure $A$ , the evolution of another observable
$B$, hence its value at the time of measurement, might be quite
different from what it would have been if $B$, and not $A$, had been measured.
As explained in item (2.6) of section (2), in our classical model, this
{\it disturbance effect\/} is not only deterministic but also known, so
that we can use it in our calculations.

For a given pair (point in the unit disk) $p\in D$, we do not know
a priori the values assumed by $S^{(j)}_x(p)$ (because $p$ is a hidden
parameter for the experimentalists). However whatever these values are
they must satisfy the inequality
$$|S^{(1)}_a(p)S^{(2)}_b(p)-S^{(1)}_c(p)S^{(2)}_b(p)|\leq1-
S^{(1)}_a(p)S^{(1)}_c(p)\eqno(1)$$
Since, in the second experiment, $S^{(2)}_b$ has been measured and
$b\not=c$, we now, from item (2.6) of section (2), that particle $2$ of the
pair $p$, has changed the value of $S^{(2)}_c$ from $S^{(2)}_c(p)$ to
$-S^{(2)}_c(p)$. Moreover, from section (4), we know that
$$S^{(2)}_c(p)=S^{(1)}_c(p)$$
Therefore, if we want to insert the value of $S^{(2)}_c(p)$ {\it that we
would have found if $S^{(2)}_c$ would have been measured instead of
$S^{(2)}_b$\/}, then we have to replace in (1), $S^{(1)}_c(p)$ by $-S^{(2)}_c(p)$
thus obtaining
$$|S^{(1)}_a(p)S^{(2)}_b(p)-S^{(1)}_c(p)S^{(2)}_b(p)|\leq1+S^{(1)}_a
(p)S^{(2)}_b(p)$$
From which Bell's inequality
$$|\langle S^{(1)}_aS^{(2)}_b\rangle-\langle S^{(1)}_c S^{(2)}_b
\rangle|\leq1+\langle S^{(1)}_aS^{(2)}_c\rangle$$
is easily deduced by taking averages.

In the following section we show that the experimental results
contradict this argument.
\bigskip

\noindent{\bff (6) Violation of the Bell's inequality}

\medskip
We will now prove that there exist three directions $a$, $b$, $c$ in the
plane such that the correlations
$$\langle S^{(1)}_aS^{(2)}_b\rangle \quad, \quad\langle S^{(1)}_cS^{(2)}_b
\rangle \quad,\quad\langle S^{(1)}_aS^{(2)}_c\rangle\eqno(1)$$
violate the Bell inequality. To this goal we fix $a$ to be the $x$--axis
and consider a generic choice of
both vectors $c,b$ in the upper semi--circle. According to our rules
the corresponding statistics is described by Figure (8) below:

\centerwmf{6cm}{7cm}{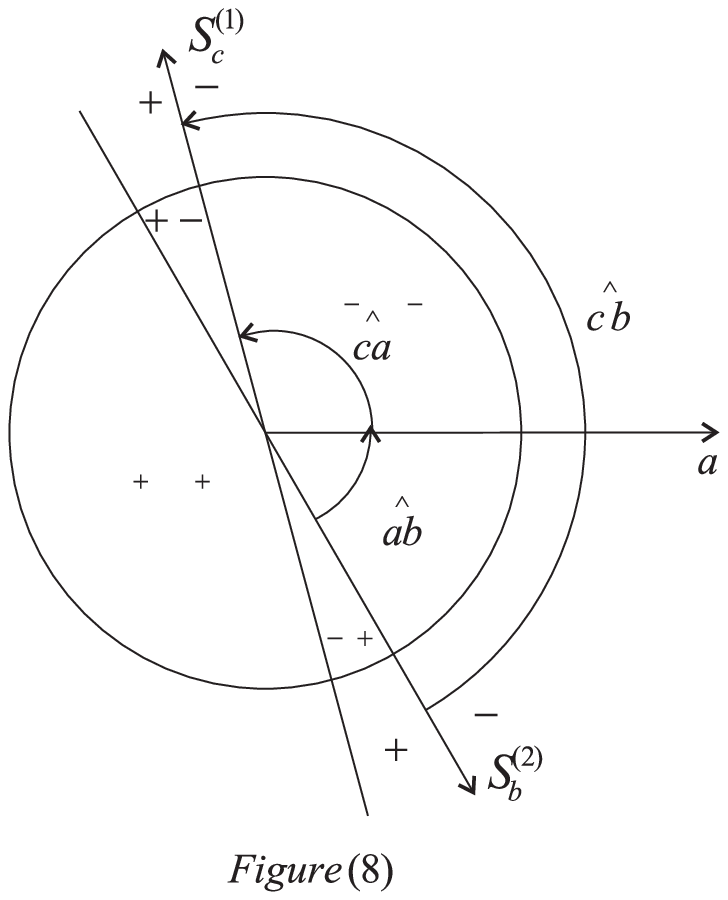}  %7.27  9.30

\vfill\eject

Denoting $(+,+)$, $(-,-)$ the probabilities of concordance, and $(-,+)$,
$(+,-)$, those of discordance, one has
$$\langle S^{(1)}_cS^{(2)}_b\rangle = 2(-,-)   - 2(+,-)=
2(-,-)-2({1\over 2} - (-,-) ) = -1 + 4(-,-)$$
Denoting $\widehat{cb}$ the probability of the $(-,-)$--concordance
for the choice $(S^{(1)}_c,S^{(2)}_b)$, this gives the general formula
$$\langle S^{(1)}_cS^{(2)}_b\rangle=-1+4\widehat{cb}\eqno(2)$$
valid for any choice of the vectors $c,b$ in the upper semi--circle.

Let us now choose the three directions $a$, $b$, $c$ in the plane as
indicated in Figure (9).
\centerwmf{11.4cm}{9.4cm}{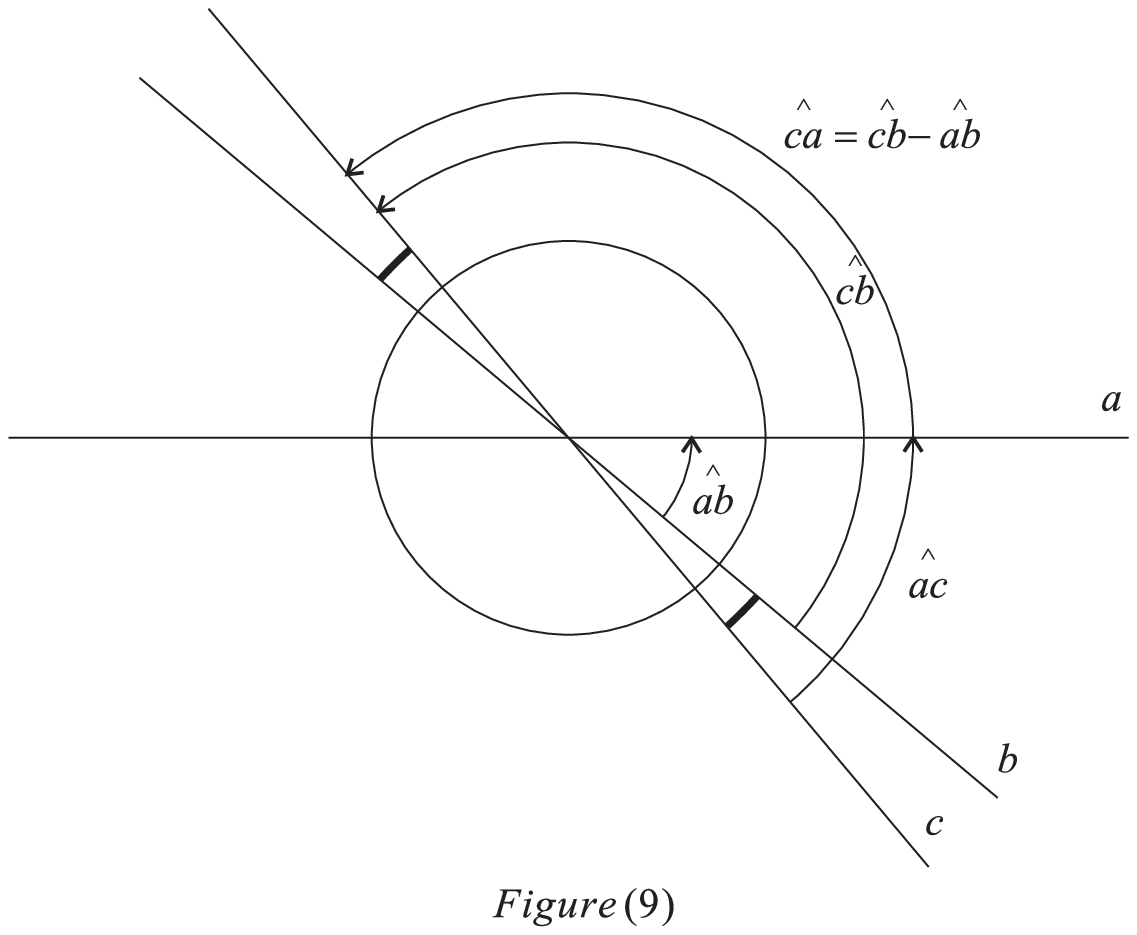}

Then, according to equation (2),
the two sides of the Bell inequality are respectively:
$$|\langle S^{(1)}_aS^{(2)}_b\rangle-\langle S^{(1)}_cS^{(2)}_b\rangle
|=4|\widehat{ab}-\widehat{cb}|=4(\widehat{cb}-\widehat{ab})$$
$$1+\langle S^{(1)}_aS^{(2)}_c\rangle=4\widehat{ac}$$
and we are reduced to compare
$$\widehat{cb}-\widehat{ab}=\widehat{ca}\qquad\hbox{and}\qquad
\widehat{ac}$$
But, because of our choice of the axes (cf. Figure 9), one has
$$\widehat{ca}>\widehat{ac}$$
Thus we conclude that
$$|\langle S^{(1)}_aS^{(2)}_b\rangle-\langle S^{(1)}_cS^{(2)}_b
\rangle|>1+\langle S^{(1)}_aS^{(2)}_c\rangle$$
which violates the Bell inequality.
\bigskip

\beginsection{\bff (7) A Greenberger--Horne--Zeilinger type contradiction}

\bigskip
Greenberger, Horne and Zeilinger [GrHoZe93] have constructed an example showing that the attempt to attribute simultaneous values to 2--valued observables represented by non commuting operators may lead to a contradiction independently of any statistical consideration.
In this section we show that our experiment also provides a classical analogue of the construction of these authors.

From (3.1) and (3.3) we know that the answers of a particle of type I to the
two (mutually exclusive) measurements $B(a)$ and $B(Ra)$ will be
opposite:
$$S^{(1)}_a=-S^{(1)}_{Ra}\eqno(1)$$
Exactly in the same sense we know that
$$S^{(2)}_a=-S^{(2)}_{Ra}\eqno(2)$$
Moreover we know that
$$S^{(1)}_a=S^{(2)}_a\eqno(3)$$
$$S^{(1)}_a=-S^{(2)}_{Ra}\eqno(4)$$
Notice that, if we interpret $S_a$ (resp. $S_{Ra}$) as the response of a
type $I$ particle to the only action of the magnetic field $B(a)$ (resp.
$B(Ra)$) the joint event $[S_a=+]$ and $[S_{Ra}=-]$ makes no sense as a
simultaneous statement on the same particle.
Now let us show that, if we pretend (as done in the original Bell's
argument and in all discussions of the EPR type experiments) that all
these relations hold
simultaneously, in the sense of joint events, then we arrive to a
contradiction. Interpreting the diagrams below as explained in section (3)
(both columns in each diagram are referred to the same particle), we see that
relations (1) and (3.1) imply that, for any vector $a$ in the upper half
plane:
$$I\to B(a)\to\cases{
S_a=+\ ;\quad S_{Ra}=-\cr
S_a=-\ ;\quad S_{Ra}=+\cr}\eqno(5)$$
and relations (2) and (3.2) imply:
$$II\to B(a)\to\cases{
S_{Ra}=+\ ;\quad S_a=-\cr
S_{Ra}=-\ ;\quad S_a=+\cr}\eqno(6)$$
Relations (1) and (3.3) imply:
$$I\to B(Ra)\to\cases{
S_{Ra}=-\ ;\quad S_a=+\cr
S_{Ra}=+\ ;\quad S_a=-\cr}\eqno(7)$$
and relations (2) and (3.4) imply:
$$II\to B(Ra)\to\cases{
S_a=-\ ;\quad S_{Ra}=+\cr
S_a=+\ ;\quad S_{Ra}=-\cr}\eqno(8)$$
But, if we pretend to attribute to $S^{(2)}_a$ the values given by the
second column of (6) then, because of
(5) this would contradict (3) because it gives
$$S^{(1)}_a=-S^{(2)}_a$$
Similarly also (7) and (8) contradict (3) because they give
$$S^{(1)}_{Ra}=-S^{(2)}_{Ra}$$
Summing up: the functions of the ``hidden parameter'' $p\in D$ correctly
describe the behavior of {\bf pairs of observables referred to different
particles}. In fact this behavior can be simulated on the computer with
arbitrary precision. However, if we pretend to extend this descriptions
to triples (or quadruples) of observables by including {\bf pairs of
observables referred to the same particle}, then we arrive to a logical
contradiction, independent of any statistics.

This kind of contradictions should not be considered as pathological but
rather as natural manifestation of the paradoxes that might arise if one
attempts to apply the conceptual schemes, elaborated having in mind a
{\it passive\/} (ballot--box like) reality, to a physical situation
involving an {\it adaptive\/} (chameleon like) reality. In the latter case
the {\it value of an observable\/} may express the dynamical reaction to an
interaction. Two incompatible interactions (e.g. {\it only magnetic field
$B(a)$\/} or {\it only magnetic field $B(Ra)$\/}), might give rise to
opposite physical reactions (e.g. {\it going up\/} or {\it going down\/}).
If we codify this reactions with numbers, e.g. $+1$ or $-1$ and then we
pretend to interpret these numbers as {\it values of observables\/}, in the
same sense of {\it passive reality\/} (e.g. color or weight of a ball in a
ballot box) then it is clear that joint events of the form $[S_a=+1]$ and
$[S_{Ra}=-1]$, when referred to the same particle at the same time, are in
trinsically contradictory. In fact the same particle cannot simultaneously
{\it go up\/} and {\it go down\/}.
\bigskip

\beginsection{\bff (8) Conclusions and acknowledgments}

\bigskip
Our experiments show that there is no contradiction between lcality,
realism and quantum theory, thus confirming the results of the theoretical
analysis of [Ac81],..., [AcRe99b].
The three directions in Figure (9)
parametrize three experiments I $(a,b)$, II $(c,b)$, III $(a,c)$, in which
the experimentalists $1$ and $2$ locally choose a unit vector according to the
scheme illustrated in Table 1.
%\vglue6.1truecm
\bigskip
$$
\matrix{
  & \hbox{I} & \hbox{II} & \hbox{III} \cr
1 & \hbox{a} & \hbox{c}  & \hbox{a}   \cr
2 & \hbox{b} & \hbox{b}  & \hbox{c}
}
$$
\centerline{Table 1}
\bigskip

Having fixed $a$ to be the $x$--axis, as in Figure (9) this is
equivalent to choose the angles $\widehat{ab}$ and $\widehat{ac}$ of
Figure 9.

For the choice of the angles illustrated in Table 2 (angles are given in
radiants, so the local choices consist in numbers between $0$ and $3,14$)
\bigskip
$$
\matrix{
  & \hbox{I} & \hbox{II} & \hbox{III} \cr
1 & 0 & 1.989675  & 0   \cr
2 & 0.3141593 & 0.3141593  & 1.989675
}
$$
\centerline{Table 2}
\bigskip

we verify that the difference between the left and the right hand side of
Bell's inequality is of order
$$0.521>0$$
leading to a violation of the inequality. Since our scheme is obviously
stable under small perturbations, we have violation of the inequality
for infinitely many directions.
\bigskip

The present experiment has been first performed on july 10, 1999 at the
Centro Volterra and first publicly performed on september 13, 1999 at the
Centre for Philosophy of Natural and Social Science of the London School of
Economics. Then, in the same year: on September 29 at the
Steklov Institute (Moscow), in occasion of the conference dedicated to the
90--th birthday of N.N.Bogoliubov; on November 9 at the 3--rd Tohwa University
International Meeting on Statistical Physics (Tohwa University, Fukuoka,
Japan); in January 10, 2000 at the IV International Workshop on Stochastic
Analysis and Mathematical Physics, in Santiago;
on February 24, 2000 at the Department of Physics of the University of
Pavia; on March 7, 2000 at the
3rd Meijo Conference on Quantum Information. It was described without
performance of the experiment (because of problems with the computer)
on December 17, 1999 at the conference on Quantum Probability and
Infinite Dimensional Analysis, Jointly organised by Indian Statistical
Institute and the Jawaharlal Nehru Centre for Advanced Scientific Research
(Bangalore, India) and on May 4, 2000  at the conference on Quantum Paradoxes
in Nottingham.

Those who want see how the experiment concretely works may consult the
Volterra Center's WEB page:  {\bf http://volterra.mat.uniroma2.it}
from where the programme can be down--loaded. In the same page one will
also find the files corresponding to the papers mentioned in the
bibliography as well as a synopsis (in English) of the book [Ac97] and other
material relevant to explain in more detail the main theses of quantum
probability.
\medskip

The authors want to express their gratitude to all those who actively
partecipated to the debate following these presentations. In particular
they are grateful to I. Volovich and G. Rizzi for many interesting
discussions and to I. Arefeva, K. Awaya, S. Kozyrev and M. Skeide, W.
von Waldenfels, for
constructive criticism which stimulated a clearer exposition of several
points in our exposition. A particular thank is due to Alessio Accardi
who pointed out the crucial role of reflections in the interpretation of
the experiment and to Mauro D'Ariano who originated the series of
experiments described in [AcRe99a], [AcRe99b] and the present one, by
suggesting that an experiment would convince physicists more than a theorem.

\bigskip

\beginsection{\bff Bibliography }

\medskip
  \item{[Ac81]} Luigi Accardi :
``Topics in quantum probability'', Phys. Rep. 77 (1981) 169-192

\item{[Ac97]}  Luigi Accardi:
Urne e camaleonti.
Dialogo sulla realt\`a, le leggi del caso e la teoria quantistica.
Il Saggiatore (1997).
Japanese translation, Maruzen (2000), russian translation, ed. by Igor
Volovich, PHASIS Publishing House (2000), english translation by Daniele
Tartaglia, to appear

\item{[Ac99]}
Luigi Accardi:
On the EPR paradox and the Bell inequality
Volterra Preprint (1998) N. 350.

\item{[AcRe99a] }
Luigi Accardi, Massimo Regoli:
Quantum probability and the interpretation of quantum mechanics: a crucial
experiment,
Invited talk at the workshop: {\it``The applications of mathematics
to the sciences of nature: critical moments and aspetcs''\/}, Arcidosso
June 28-July 1 (1999). To appear in the proceedings of the workshop, Preprint
Volterra N. 399

\item{[AcRe99b]}
Luigi Accardi, Massimo Regoli:
Local realistic violation of Bell's inequality: an experiment,
Conference given by the first--named author at the Dipartimento di Fisica,
Universit\`a di Pavia on 24-02-2000, Preprint Volterra N. 402

\item{[Be64]}
J.S. Bell:
On the Einstein Podolsky Rosen Paradox
Physics 1 no. 3  (1964) 195-200

\item{[Be87]}
J.S. Bell: Speakable and Unspeakable in Quantum Mechanics,
Cambridge, Cambridge University Press (1987)

\item{[CuMcM89]}
J.T. Cushing, E. McMullin (eds.):
Philosophical Consequences of Quantum Theory
Reflections on Bell' s Theorem.
University of Notre Dame Press, vol. II (1989)

\item{[EPR35]}
A. Einstein, B. Podolsky, N. Rosen:
Can quantum mechanical description of
reality be considered complete?
Phys. Rev. 47(1935)777-780

\item{[GrHoZe93] }
A. Zeilinger, M.A. Horne, D.M. Greenberger:
Multiparticle interferometry and the superposition principle.
American Institute of Physics (1993)

\item{[Maud94]}
Tim Maudlin:
Quantum Non--locality and relativity, Blackwell (1994)

\item{[Mdx88]}
J. Maddox:
New ways with Bell's inequalities.
Nature 335 (1988)

\item{[PrTe93]}
Press H. William, A. Teukolsky Saul:
Numerical recepees in $C$. The art of scientific computing,
Cambridge University Press, 1993.

\item{[Red87]}
M. Redhead:
Incompleteness, nonlocality and realism,
Clarendon Press (1987)

\item{[Sak85]}
Sakurai Jun John:
Modern Quantum Mechanics, Benjamin (1985)
\bye